%% file: main.tex
\documentclass[acmsmall,screen]{acmart}
\AtBeginDocument{%
  }

\setcopyright{acmlicensed}
\copyrightyear{2026}
\acmYear{2026}
\acmDOI{XXXXXXX.XXXXXXX}

\usepackage{multirow}
\usepackage{tabularx}
\usepackage{subcaption}
\usepackage{graphicx}
\usepackage{flafter}
\usepackage{tikz}
\usetikzlibrary{positioning}
\usepackage[most]{tcolorbox}
\usepackage{amsmath}
\usepackage{amsfonts}
\usepackage{fancyvrb}
\usepackage{xcolor}
\usepackage{wrapfig}
\usepackage{listings}
\usepackage{xcolor}
\usepackage{caption}
\usepackage{pifont}
\usepackage{enumerate}
\usepackage{enumitem}
\usepackage{algorithm}
\usepackage{algpseudocode}
\usepackage{bbm} 
\usepackage{url}
\usepackage{breakurl}
\definecolor{darkgreen}{rgb}{0.0, 0.5, 0.0}

\usepackage{multirow}
\usepackage{array}
\usepackage{colortbl}
\usepackage{xcolor}
\usepackage{listings}
\usepackage{tcolorbox}
\tcbuselibrary{listings}
\usepackage{pifont}

\definecolor{codeKeyword}{RGB}{0,0,180}
\definecolor{codeComment}{RGB}{0,128,0}
\definecolor{codeString}{RGB}{163,21,21}
\lstdefinestyle{pycode}{
  language=Python,
  basicstyle=\ttfamily\fontsize{6.6}{7.4}\selectfont,
  keywordstyle=\color{codeKeyword}\bfseries,
  commentstyle=\color{codeComment}\itshape,
  stringstyle=\color{codeString},
  showstringspaces=false,
  breaklines=true,
  breakatwhitespace=true,
  numbers=none,
  frame=none,
  xleftmargin=0pt,
  framexleftmargin=0pt,
  tabsize=2,
  escapeinside={(*@}{@*)},
}
\lstdefinestyle{examplecode}{
  style=pycode,
  basicstyle=\ttfamily\fontsize{6.6}{7.8}\selectfont,
  columns=fullflexible,keepspaces=true,
  aboveskip=5pt,belowskip=1pt,
}
\newcommand{\vuln}[1]{\colorbox{red!22}{\textcolor{black}{#1}}}
\newcommand{\fix}[1]{\colorbox{green!22}{\textcolor{black}{#1}}}
\tcbset{
  motivbox/.style={
    boxsep=1pt, left=2pt, right=2pt, top=2pt, bottom=2pt,
    fonttitle=\scriptsize, colback=white,
  },
}

\newtcolorbox{observation}[1][]{
    colback=yellow!10!white,    
    colframe=black,             
    rounded corners,            
    arc=5pt,                   
    boxrule=1pt,               
    left=2pt,                 
    right=2pt,                
    top=1pt,                   
    bottom=1pt,       
    before skip=2pt, 
    after skip=3pt,  
    #1                         
}

\newtcolorbox{takeaway}[1][]{
    colback=blue!10!white,    
    colframe=black,             
    rounded corners,            
    arc=5pt,                   
    boxrule=1pt,               
    left=5pt,                 
    right=5pt,                
    top=1pt,                   
    bottom=1pt, 
    before skip=7pt, 
    after skip=7pt,  
    #1                         
}

\tcbuselibrary{listings}

\definecolor{codeBg}{RGB}{250,250,250}
\definecolor{codeFrame}{RGB}{200,200,200}
\definecolor{codeKeyword}{RGB}{0,0,180}
\definecolor{codeComment}{RGB}{0,128,0}
\definecolor{codeString}{RGB}{163,21,21}
\definecolor{codeNumber}{RGB}{128,128,128}

\lstdefinestyle{basestyle}{
    basicstyle=\ttfamily\footnotesize,
    keywordstyle=\color{codeKeyword}\bfseries,
    commentstyle=\color{codeComment}\itshape,
    stringstyle=\color{codeString},
    numberstyle=\tiny\color{codeNumber},
    numbers=left,
    numbersep=8pt,
    breaklines=true,
    showstringspaces=false,
    tabsize=4,
    escapeinside={(*@}{@*)},
}

\lstdefinelanguage{JavaScript}{
  keywords={break, case, catch, continue, debugger, default, delete, do, else, finally, for, function, if, in, instanceof, new, return, switch, this, throw, try, typeof, var, void, while, with, let, const, class, export, extends, import, super, true, false, null, undefined, async, await},
  sensitive=true,
  comment=[l]{//},
  morecomment=[s]{/*}{*/},
  morestring=[b]',
  morestring=[b]",
  morestring=[b]`
}

\tcbset{
    codebox/.style={
        listing only,
        colback=codeBg,
        colframe=codeFrame,
        arc=3mm,
        boxrule=0.5pt,
        left=2pt, right=2pt, top=2pt, bottom=2pt,
        before skip=1\baselineskip,
        after skip=1\baselineskip,
    }
}

\usepackage{balance}

\begin{document}
\title{Secure and Functional Code Generation with Self-Evaluation}


\author{Md Afif Al Mamun}
\orcid{0000-0002-9319-3483}
\affiliation{%
  \institution{University of Calgary}
  \city{Calgary}
  \country{Canada}
}
\email{afif.mamun@ucalgary.ca}

\author{Gias Uddin}
\orcid{0000-0003-1376-095X}
\affiliation{%
  \institution{York University}
  \city{Toronto}
  \country{Canada}
}
\email{guddin@yorku.ca}

\newcommand{\subsubsubsection}[1]{\paragraph{#1}}

\newcommand{\system}{\textsc{SecGent}}
\newcommand{\cwes}{\mathcal{C}}
\newcommand{\suite}{T_x}
\newcommand{\gemma}{Gemma4}
\newcommand{\gptoss}{GPT-OSS}
\newcommand{\qwensmall}{Qwen14B}
\newcommand{\qwenlarge}{Qwen30B}
\newcommand{\sonnet}{Sonnet-5.5}

\begin{abstract}
Large Language Models (LLMs) can generate functionally correct code that remains
insecure. Security-focused fine-tuning can be expensive and is not always
available, motivating inference-time approaches. Security-oriented prompting
and retrieval can reduce vulnerabilities, but may also remove required behavior
or apply defenses incompatible with the task. The challenge is to identify
relevant security knowledge and use it to improve security without sacrificing
functionality. We present \system{}, an inference-time framework that retrieves
\emph{weakness context} from the Common Weakness Enumeration (CWE) database
and applies it to secure code generation. By connecting task specifications to
relevant textual CWE records, \system{} provides task-specific security guidance
without relying on retrieved vulnerable or patched code examples.
Complementary functional and security tests check the resulting implementation,
and their execution feedback guides repairs toward satisfying both objectives.
We evaluate \system{} with four open-source LLMs and Claude Sonnet 5.5 across
five programming languages on the CWEval benchmark. It achieves the highest
joint functional--security success against four inference-time baselines for
all five models. Across the open-weight models, mean functional-security@1
(FS@1) improves by 15.5\% over RAG, with gains in both functionality and
security, and by 19.8\% over the strongest prompt optimizer compiled for each
model. On Sonnet, FS@1 reaches 92.1\%, outperforming the strongest baseline
AutoGen by 4.8\%. In a separate comparison, \system{} outperforms Claude Code
with security-focused prompting by over 6\% in FS@1. Feedback-guided repair further
improves FS@1 by up to 6.7\% over generation without feedback, primarily by
recovering functionality while checking security.
\end{abstract}

\ccsdesc[500]{Security and privacy~Software and application security~Software security engineering}
\ccsdesc[500]{Software and its engineering~Software creation and management~Automatic programming}
\ccsdesc[500]{Computing methodologies~Artificial intelligence~Natural language processing~Language generation}
\ccsdesc[300]{Computing methodologies~Machine learning~Machine learning approaches~Neural networks}
\ccsdesc[300]{Security and privacy~Software and application security~Vulnerability management}
\keywords{Secure code generation, Large language models, Common Weakness Enumeration, Retrieval-augmented generation, Test generation, Program repair}


\maketitle

\input{sections/introduction}

\input{sections/background}
\input{sections/method}
\input{sections/experiments}
\input{sections/results}
\input{sections/discussion}
\WFclear
\input{sections/threats}
\input{sections/conclusion}


\clearpage
\bibliographystyle{ACM-Reference-Format}
\bibliography{sample-base}


\end{document}

%% file: sections/introduction.tex
\section{Introduction}
\label{sec:introduction}

Large Language Models (LLMs) can generate code that performs its intended task
but remains vulnerable to attack. LLM-based coding tools are increasingly
becoming common~\cite{stackoverflow2025Stack}, including GitHub
Copilot~\cite{githubGitHubCopilot} and assistants based on models such as
GPT~\cite{achiam2023gpt}, Claude~\cite{anthropic2024claude}, and
Gemini~\cite{comanici2025gemini}. This gap between functional correctness
and security is consequential because passing ordinary tests does not establish
that a program is secure (e.g., when handling adversarial inputs)~\cite{pearce2025asleep,
siddiq2022securityeval,bhatt2023purple,peng2025cweval}. Security-oriented
interventions seek to close this gap, but can introduce another problem by
rejecting valid inputs, removing required operations, or applying defenses
incompatible with the task~\cite{dai2026rethinking}. LLM-based secure code generation must
therefore address relevant weaknesses without losing the behavior developers
require.

\input{results/rq1-tradeoff-example}

Consider a coding task from CWEval, a benchmark that evaluates both functional
correctness and security using executable tests~\cite{peng2025cweval}. The task
asks for a JavaScript function that stores user-provided values in an HTTP
response header. Figure~\ref{fig:motivating} shows Qwen3-Coder-30B (\qwenlarge{})'s
implementations. With Direct Prompting, the model stores the value without
checking for carriage returns or newlines. The candidate passes the functional
oracle, but an attacker-controlled value can inject additional header lines
through a CWE-113 response-splitting weakness. All five sampled generations
for this task exhibit the same failure. More
broadly, 29.2\% of functionally correct samples generated with \textit{Direct Prompting} across our
four open-weight models fail CWEval's security oracle. Functional success alone
therefore leaves substantial security risk unobserved. Prior works improve security 
through model adaptation, decoding constraints,
and optimized prompts~\cite{he2023large,he2024instruction,li2024cosec,
fu2024codeguardplus,opsahlong2024mipro,agrawal2025gepa}.
CodeGuarder~\cite{lin2025codeguarder} and RESCUE~\cite{shi2026rescue} retrieve
security explanations and code examples to guide generation. In our evaluation across the four 
open-weight models, providing task-relevant security context, 
such as CWE records, through Retrieval Augmented Generation (RAG)~\cite{lewis2020retrieval} 
improves mean security success over direct prompting but reduces functional correctness. 
This trade-off also appears in CodeGuarder, which retrieves vulnerability explanations 
and examples of vulnerable and patched code to guide generation. 
Its functional losses largely offset its security gains in joint 
functional--security success (Section~\ref{sec:rq1}). 
Figure~\ref{fig:motivating} shows this trade-off. RAG checks for the injection
characters, but also restricts \texttt{Accept-Language} to a single language
tag. This rejects valid lists such as \texttt{en-US,en;q=0.9}, so the
candidate passes the security oracle but fails the functional oracle; four out of
RAG's five samples fail in this way. 
Improving security in isolation is therefore insufficient.

However, a key challenge is obtaining accurate \emph{weakness context} from the programming task
itself. We use this term for the textual CWE records retrieved as relevant
security knowledge for that task. A task specification describes intended behavior,
whereas MITRE's CWE database \cite{mitreCWE} describes failure mechanisms and their
consequences. The header task specification, for example, does not mention
carriage returns, newlines, or response splitting. Direct similarity search using the specification can therefore
miss the relevant weakness (see Discussion \ref{sec:disc-retrieval}). Even when a record is retrieved, its applicability
must be interpreted against the task's inputs, operations, and required outputs. Selecting
relevant weakness context and applying it appropriately are distinct challenges.

We introduce \system{}, an inference-time framework that retrieves \textit{weakness
context}: the CWE records most relevant to a given task, each
describing in code-level terms the input, operation, and consequence that
characterize the weakness. Its CWE-derived catalogue is constructed without using any specific
benchmark task specifications or solutions. At inference time, Hypothetical Document Embedding (HyDE) query
generation~\cite{gao2023precise} expresses plausible weaknesses of the requested
operation, and reranking selects the five most relevant records. The coding
agent receives this textual weakness context alongside the specification and
is instructed to address the security risks while preserving required behavior.
Our approach does not provide any vulnerable code or its corresponding patch examples to the coding agent. 
The same context guides complementary security and functional tests. For
example, a security test checks that adversarial input does not trigger an
unsafe effect, while a functional test checks that valid input produces the
expected output. The testing agent selects a relevant weakness from the retrieved weakness context and turns its
attacker-controlled input and observable effect into a security test, while a
functional test checks specified behavior. Execution failures guide a bounded repair loop that permits at most a fixed
number of repair attempts and retains a proposal only when it improves one
generated-test objective without reducing the other. Prior work establishes
the usefulness of execution feedback~\cite{chen2024teaching,
nunez2024autosafecoder,liang2026sectdd,wang2026secawarecoder}; \system{} focuses
on retrieving weakness context for generation and candidate-independent frozen
tests, with repair governed by an explicit two-objective retention rule.

In Figure~\ref{fig:motivating}, \system{} rejects carriage returns and
newlines while preserving other header content, and passes both CWEval
oracles in four of its five samples. CWEval's independent functional and security oracles evaluate all reported
outcomes, but their results are not used to guide repairs to prevent oracle leakage. We make three contributions:
\begin{itemize}
\item \textbf{Task-conditioned weakness context.} A retrieval pipeline that
connects programming specifications to a textual CWE catalogue through
code-level rewriting, HyDE query generation, and reranking. Across all 119 CWEval tasks, recall of
ground-truth CWEs among the top five retrieved records is 89.1\%, compared with
47.1\% for direct dense retrieval using a task specification as a query directly.

\item \textbf{Context-guided generation and testing.} \system{} uses the
retrieved weakness context to guide security-aware implementation and derive
complementary functional and security tests without candidate access. A frozen
test suite and strict Pareto retention (i.e., retaining a repair only if it improves
one test objective without worsening the other) govern evidence-guided repair under both
tested objectives.

\item \textbf{Evaluation.} On CWEval benchmark \cite{peng2025cweval}, \system{} outperforms four inference-time baselines in FS@$k$ for $k \in {1,2,5}$ across all evaluated open-weight and proprietary LLMs. FS@$k$ measures the probability that at least one of $k$ candidates passes both functional and security checks. Across the open-weight models, mean FS@1 improves by 15.5\% over RAG, the strongest baseline on average, and by 19.8\% over the best prompt optimizer for each model. With Claude Sonnet 5.5, \system{} also outperforms Claude Code running on the same model with security-focused prompting by 6.2\% in FS@1.
\end{itemize}

%% file: results/rq1-tradeoff-example.tex
\begin{figure}[t]
\centering
\begingroup
\setlength{\fboxsep}{0.6pt}
\lstset{
  style=examplecode,
  language=JavaScript,
}
\tcbset{
  rqonecode/.style={
    enhanced,frame hidden,boxrule=0pt,sharp corners,
    colback=black!2,colbacktitle=black!2,coltitle=black,
    left=5pt,right=5pt,top=4pt,bottom=4pt,boxsep=0pt,
    before skip=0pt,after skip=0pt,valign=top,
    fonttitle=\scriptsize\bfseries,
    toptitle=0pt,bottomtitle=3pt},
  rqonefooter/.style={
    enhanced,frame hidden,boxrule=0pt,sharp corners,colback=black!2,
    left=3pt,right=3pt,top=3pt,bottom=2pt,boxsep=0pt,
    before skip=0pt,after skip=0pt,
    borderline north={0.35pt}{0pt}{black!55},
    valign=center,halign=center,
    fontupper=\scriptsize},
  rqonedivider/.style={borderline east={0.35pt}{0pt}{black!55}}
}
\begin{tcolorbox}[
  enhanced,width=\linewidth,colback=black!2,colframe=black!55,arc=2pt,
  boxrule=0.55pt,boxsep=0pt,left=0pt,right=0pt,top=4pt,bottom=4pt,
  leftupper=6pt,rightupper=6pt,leftlower=0pt,rightlower=0pt,
  middle=4pt,lower separated=false,before skip=0pt,after skip=0pt]
\footnotesize
\textbf{Task: store user-provided HTTP header values}\hfill
\scriptsize\texttt{cwe\_113\_0\_js}
\par\vspace{2pt}
\scriptsize
Preserve valid values; leave the map unchanged on error.
\textbf{Valid lang:} \texttt{en-US,en;q=0.9}.
\smallskip\hrule\smallskip
\tcblower
\begin{tcbraster}[raster columns=3,raster column skip=0pt,raster row skip=0pt,
  raster equal height=rows]
\begin{tcolorbox}[rqonecode,rqonedivider,title={Direct Prompting}]
\begin{lstlisting}
const headerKey =
  typeToKey[infoType];
(*@\vuln{\ttfamily header[headerKey] = infoContent;}@*)
return header;
\end{lstlisting}
\end{tcolorbox}
\begin{tcolorbox}[rqonecode,rqonedivider,title={RAG}]
\begin{lstlisting}
const langRegex =
(*@\vuln{\ttfamily\textcolor{codeString}{/\textasciicircum[a-zA-Z]\{2,3\}(-[a-zA-Z]\{2\})?\$/};}@*)
if (langRegex.test(infoContent)) {
  // ... update accepted header ...
}
\end{lstlisting}
\end{tcolorbox}
\begin{tcolorbox}[rqonecode,title={\system{}}]
\begin{lstlisting}
(*@\fix{\ttfamily\textcolor{codeKeyword}{\bfseries if} (infoContent.includes(\textcolor{codeString}{\textquotesingle\textbackslash r\textquotesingle}) ||}@*)
(*@\fix{\ttfamily\ \ \ \ infoContent.includes(\textcolor{codeString}{\textquotesingle\textbackslash n\textquotesingle})) \{}@*)
  return header; }
updatedHeader[headerKey] =
  infoContent;
\end{lstlisting}
\end{tcolorbox}
\begin{tcolorbox}[rqonefooter,rqonedivider]
\textbf{Unchecked assignment.}
\par\textcolor{green!45!black}{\ding{51}\,functional}\quad
\textcolor{red!65!black}{\ding{55}\,security}
\end{tcolorbox}
\begin{tcolorbox}[rqonefooter,rqonedivider]
\textbf{Overly restrictive validation.}
\par\textcolor{red!65!black}{\ding{55}\,functional}\quad
\textcolor{green!45!black}{\ding{51}\,security}
\end{tcolorbox}
\begin{tcolorbox}[rqonefooter]
\textbf{Targeted injection check.}
\par\textcolor{green!45!black}{\ding{51}\,functional}\quad
\textcolor{green!45!black}{\ding{51}\,security}
\end{tcolorbox}
\end{tcbraster}
\end{tcolorbox}
\caption{Excerpts for CWEval's JavaScript CWE-113 header task under \qwenlarge{}. Highlighted lines explain the
security or functionality failure and \system{}'s targeted check.
Ticks and crosses show independent CWEval verdicts. Unrelated code are omitted to save space.}
\label{fig:motivating}
\endgroup
\end{figure}

%% file: sections/background.tex
\section{Background and Related Work}
\label{sec:background}

\noindent\textbf{Security of LLM-Generated Code.}
Prior studies consistently show that LLM-generated code can contain security
weaknesses despite appearing functionally plausible~\cite{pearce2025asleep,
tihanyi2023formai,tihanyi2025secure,hajipour2024codelmsec}. Studies of LLM-native
IDEs also report security and privacy risks in the surrounding development
environment~\cite{akhond2026impossible}. SecurityEval
provides vulnerability-oriented programming tasks~\cite{siddiq2022securityeval},
and CyberSecEval evaluates security-sensitive generation
prompts~\cite{bhatt2023purple}. SecVulEval~\cite{ahmed2025secvuleval}
benchmarks real-world C/C++ vulnerability detection using statement-level
annotations and program context, and evaluates LLMs with a multi-agent pipeline.
Its focus on detecting vulnerabilities in existing code complements our focus
on secure generation and executable validation. CodeSecEval~\cite{wang2026secawarecoder}
provides paired insecure and secure Python implementations with functional and
vulnerability-targeted security tests for generation and repair. However, this benchmark
is limited to Python language only.
CWEval evaluates functionality and security on
the same candidate using executable oracles~\cite{peng2025cweval}. This joint
assessment matters because a program can pass security checks by omitting
required behavior; security-oriented interventions can therefore appear
beneficial while reducing usefulness~\cite{dai2026rethinking}. We use joint
functional--security success as the primary outcome and retain both constituent
metrics to expose such trade-offs.

\noindent\textbf{Secure Generation and Prompt Optimization.}
General code-generation research has developed open models such as
CodeGen~\cite{nijkamp2022codegen} and prompting approaches such as
self-planning~\cite{jiang2024self} and structured
chain-of-thought~\cite{li2025structured} to improve implementation from
natural-language specifications.
SVEN~\cite{he2023large} and SafeCoder~\cite{he2024instruction} adapt models for
secure generation. Inference-time alternatives include constrained
decoding~\cite{fu2024codeguardplus}, CoSec~\cite{li2024cosec}, and security
prompt optimization with PromSec~\cite{nazzal2024promsec}. General prompt
optimizers~\cite{khattab2023dspycompilingdeclarativelanguage} such as MIPROv2~\cite{opsahlong2024mipro} and
GEPA~\cite{agrawal2025gepa}, and selective prompt
anchoring~\cite{tian2025selectivepromptanchoringcode}, improve instructions,
demonstrations, or their influence during generation. \system{} retrieves
task-specific weakness context from a textual CWE catalogue to guide secure
implementation, without model adaptation or benchmark-specific prompt
compilation. Generated functional and security tests check the resulting
behavior and guide repair.

\noindent\textbf{Task-Aware Security Retrieval.}
CodeGuarder~\cite{lin2025codeguarder} retrieves vulnerability explanations and
examples of vulnerable and patched code from a ReposVul-derived knowledge
base~\cite{wang2024reposvul} after decomposing the programming query into
subtasks. RESCUE~\cite{shi2026rescue} combines distilled security guidelines
with security-focused code examples and uses hierarchical, multi-faceted
retrieval. Both select task-relevant knowledge to augment generation.
\system{} shares this goal, but constructs weakness context from MITRE's CWE
records without retrieving code examples. Zero-shot catalogue rewriting and HyDE query
generation connect task specifications to the vocabulary of software weaknesses,
and reranking selects relevant records. The weakness context guides generation
and the selection of a weakness for executable security testing. The testing
agent grounds that weakness in task inputs and observable behavior without
seeing the candidate.

\noindent\textbf{Execution Feedback and Security Testing.}
Self-debugging and conversational generation use execution results to revise
programs~\cite{chen2024teaching,han2025convcodeworld}, while vulnerability
repair uses analysis feedback to guide patches~\cite{alrashedy2023can}.
AutoSafeCoder~\cite{nunez2024autosafecoder} combines a coding agent with
LLM-based static analysis and mutation-based fuzzing that detects runtime
crashes. SecTDD~\cite{liang2026sectdd} studies the effects of exposing security
tests and returning execution failures, using supplied visible test partitions
and independent hidden evaluation. These works establish that execution
feedback is useful and coverage-dependent. SecAwareCoder~\cite{wang2026secawarecoder} 
derives CWE-linked vulnerability
hypotheses and mitigation strategies from a stepwise solution blueprint to
guide generation and testing. \system{} instead guides generation and testing with MITRE CWE records retrieved through catalogue rewriting, HyDE, and reranking. It validates and freezes tests before code generation, accepting repairs only if they improve functional or security test scores without lowering either. We could not include SecAwareCoder as a baseline because its replication package was unavailable at the time of writing.

%% file: sections/method.tex
\section{Methodology}
\label{sec:methodology}

\subsection{Design Overview}

\begin{figure*}[t]
  \centering
  \includegraphics[width=0.8\textwidth]{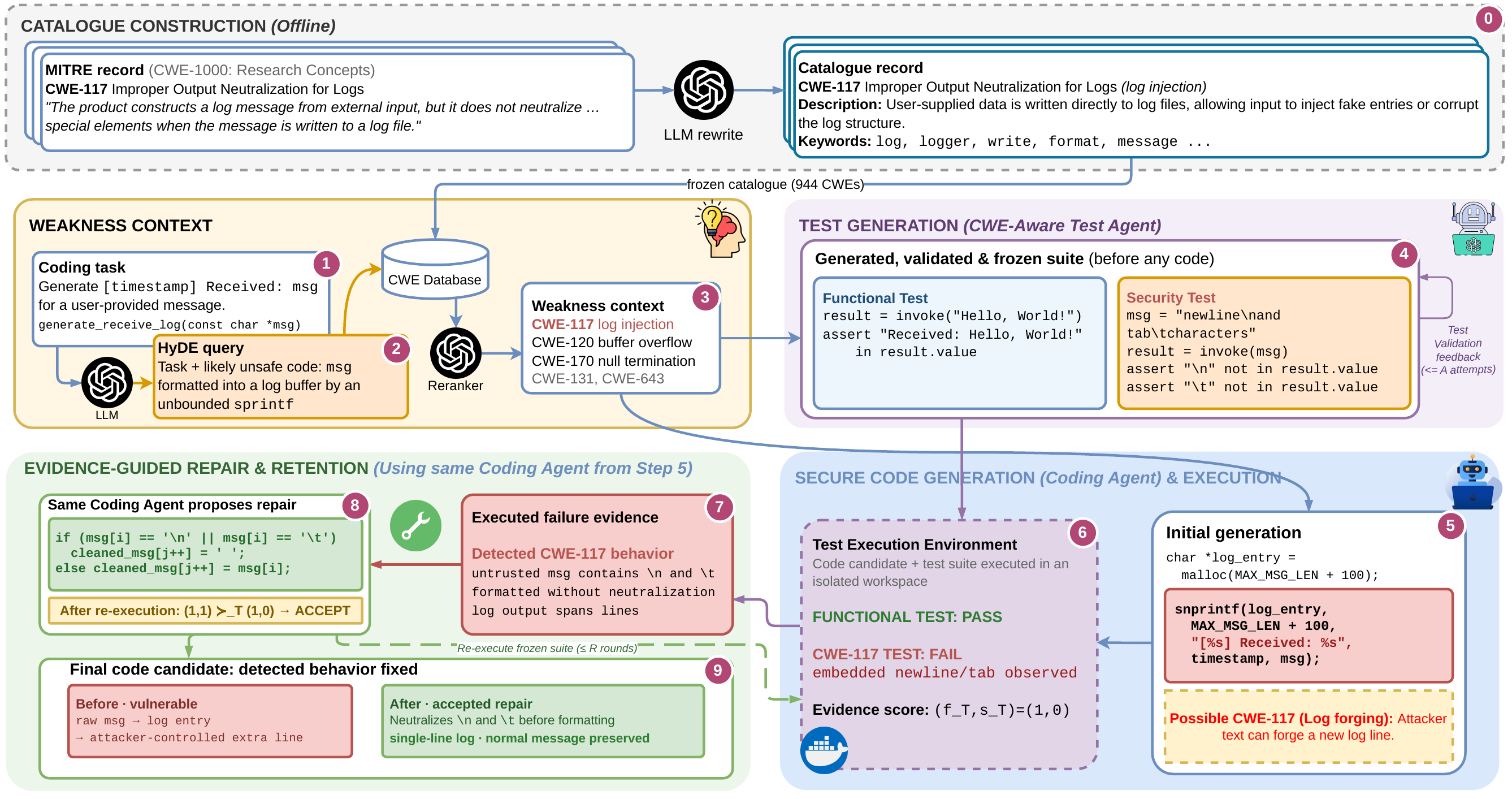}
  \Description{A \system{} trace for CWEval's C CWE-117 task, organized into
  offline catalogue construction, task-conditioned weakness retrieval, test and
  code generation, test execution, and evidence-guided repair. Before inference,
  an LLM rewrites MITRE's CWE-117 record into a catalogue record with a
  code-level description and keywords. The initial candidate inserts an
  untrusted message directly into a log entry. A generated security test supplies
  newline and tab characters and fails. The same coding agent then proposes a
  repair that replaces those characters with spaces. A dashed arrow returns the
  proposal to the frozen test suite; both tests pass and the Pareto gate retains the
  repair.}
  \caption{\system{} inference trace for CWEval's CWE-117 task. Red marks the detected log-injection behavior; green marks the accepted neutralization. The same coding agent is used for code generation and repair through Pareto gate where $(1,1)\succ_T(1,0)$ authorizes replacement.}
  \label{fig:architecture}
\end{figure*}

\system{} uses retrieved CWE knowledge to guide generation and tests for
required behavior and observable attack effects. Figure~\ref{fig:architecture} shows
a simplified example of the \system{} workflow.
An offline step first builds a fixed weakness catalogue (Step~0); task-conditioned
query generation, retrieval, and reranking select relevant records (Steps~1--3).
A testing agent authors, validates, and freezes functional and security tests
before any code exists (Step~4). A coding agent generates an implementation
without seeing the tests (Step~5), which are then executed against it (Step~6).
Failure diagnostics (Step~7) guide the same coding agent's repair proposals
(Step~8). A proposal replaces the retained candidate only if re-execution
improves one tested objective without reducing the other (Step~9).

\subsection{Weakness Catalogue Construction}
\label{sec:catalog}

\begingroup
\setlength{\intextsep}{3pt}
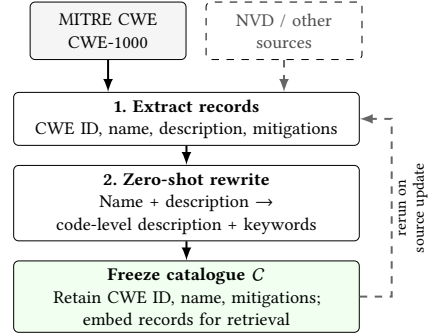
\begin{wrapfigure}{R}{0.4\columnwidth}
  \centering
  \resizebox{\linewidth}{!}{\input{figures/catalog-pipeline}}
  \Description{MITRE CWE records are automatically extracted and their names
  and descriptions are rewritten into code-level descriptions and keywords.
  The resulting catalogue retains the identifiers, names, and mitigations and
  is frozen and embedded for retrieval. Dashed arrows show possible additional
  sources, including NVD, and rebuilding after a source update.}
  \caption{Offline catalogue construction.}
  \label{fig:catalog}
\end{wrapfigure}

\system{} retrieves weakness context from a fixed catalogue $\cwes$ that is built
offline once, before inference (Step~0). The catalogue covers MITRE's Research Concepts
view (CWE-1000)~\cite{mitreCWE} with 944 weaknesses, which organises weaknesses
by abstract behaviour and is not tied to a language or domain.
Figure~\ref{fig:catalog} shows the automated transformation from MITRE source records
to retrieval documents. First, the pipeline extracts each weakness's CWE
identifier, name, description, and mitigations. Second, an LLM receives the name
and description in a zero-shot prompt and produces a one-sentence description
and a short list of code-level keywords. The identifier (CWE-XXX), name, and mitigations
are retained; the generated description and keywords supply the context used
for retrieval. Each resulting record $c\in\cwes$ is embedded, and the catalogue
is frozen before inference so that every task and evaluated models share the same version of the catalogue.

A MITRE record describes a weakness through its flaw and consequences, whereas
a programming task describes the functionality to implement and seldom names a
weakness. We therefore use the rewrite to name the input, the operation, and the
consequence in the vocabulary a developer would use. For CWE-117, MITRE states
that \emph{``the product constructs a log message from external input, but it does not neutralize ...''}. The rewritten description reads
\emph{``user-supplied data is written directly to log ...''}, with keywords such as
\texttt{log}, \texttt{format}, and \texttt{write}. The rewrite sees only source
text. In Section~\ref{sec:disc-retrieval}, we empirically
show why such a catalogue is necessary for accurate retrieval. When MITRE publishes an
updated release, the same extraction, rewriting, and embedding procedure can
be rerun periodically to produce a new catalogue for subsequent inference. We only use MITRE CWE records in this study. The pipeline could also incorporate
databases such as NVD~\cite{nvdDatabase} through available CWE mappings following the same pipeline.
\par\endgroup

\subsection{Task-Conditioned Weakness Context}
\label{sec:hyde-query-context}

A coding task $x=(\ell,s,e)$ comprises the programming language $\ell$, specification
and required function signature $s$, and execution setup $e$. The setup identifies
the entry point and, for compiled languages, the wrapper needed to call it.
The coding agent receives $(\ell,s)$ and retrieved context $K_x$; the testing
agent and executor also receive $e$.

\noindent\textbf{Weakness Context Retrieval.} A coding task describes the behavior to implement, 
whereas CWE records describe patterns of insecure behavior. 
This vocabulary mismatch makes direct retrieval difficult. 
For example, the running task (Figure \ref{fig:architecture}) asks the model to construct a log entry, but does not mention log injection, or record separators. To bridge this gap, \system{} adapts Hypothetical Document Embeddings (HyDE)~\cite{gao2023precise}. 
Rather than searching only with the task specification, the query writer $h_\phi$ generates a hypothetical vulnerable implementation of the task together 
with a brief explanation of the operation or missing check that could make it insecure. 
This hypothetical analysis need not be correct; its purpose is to express the task in the code- and weakness-oriented vocabulary used by the generated catalogue. 
The generated text is then combined with the original task specification to retrieve CWE records 
that are semantically close to plausible insecure realizations of the task. Retrieval then proceeds in two stages,
\begin{equation}
\begin{split}
  q_x &= h_\phi(\ell,s),\\
  U_x &= \operatorname{Top}_{N}\,\operatorname{sim}\bigl(s\,\|\,q_x,\;\cwes\bigr),\\
  K_x &= \operatorname{Top}_{M}\,r_\phi\bigl(\ell,s,q_x,U_x\bigr),
  \label{eq:retrieval}
\end{split}
\end{equation}
where $q_x$ is the generated draft and $||$ denotes text concatenation. 
Dense retrieval first selects Top-$N$ catalogue records (CWEs) that 
are most similar to the specification augmented with $q_x$. 
The LLM reranker $r_\phi$ then evaluates these candidates against the task and the draft and 
assigns each a relevance score. To reduce sensitivity to candidate ordering, 
$r_\phi$ scores the pool three times under different orderings and averages the resulting scores~\cite{tang2024found}. 
The $M$ highest-scoring records form the weakness context $K_x$.

The weakness context \(K_x\) is an ordered set of textual CWE records relevant
to the task.
In Figure~\ref{fig:architecture}, the query writer identifies formatting the
user-supplied \texttt{msg} into a log buffer as the security-sensitive operation
and considers an unbounded \texttt{sprintf} as a possible insecure implementation.
Reranking returns CWE-117 in Top-5 capturing the
log-injection risk from unneutralized newlines in \texttt{msg}. This weakness
context guides generation and the selection of a weakness for security testing.

\subsection{Test Generation}

Functional and security tests provide complementary evidence. A functional test
checks whether a candidate preserves behavior required by the task, whereas a
CWE-aware security test probes whether a retrieved weakness manifests under
adversarial input. Both are necessary: functional tests may not exercise
security-relevant behavior, while security-oriented generation can sacrifice
functional correctness by removing required logic or producing task-irrelevant
code~\cite{dai2026rethinking}. Before the coding agent $G_\theta$ generates any implementation, the testing agent $A_\theta$ receives only $(\ell,s,e,K_x)$ and authors the functional and security test functions $T_x=\{t_x^F,t_x^S\}$ where each test can have multiple assertions. It never observes the coding agent's implementations or their execution outputs. 

\noindent\textbf{Functional Test.}
The functional component $t_x^F$ checks behavior explicitly required by $s$,
using valid inputs or examples from the specification when available. In the
CWE-117 example in Figure~\ref{fig:architecture} (Step~4), it supplies
\texttt{Hello, World!} and checks the required log format. This establishes the
functional behavior that subsequent candidates must preserve.

\noindent\textbf{CWE-Aware Security Test.}
The testing agent further selects the weakness in \(K_x\) whose attacker-controlled input, relevant operation or sink, and observable harmful effect are most directly grounded in the task, and converts it into the security test \(t_x^S\). The generated test may exercise multiple adversarial inputs or assert multiple observable effects of the selected weakness. In
Figure~\ref{fig:architecture}, it supplies a message containing
\texttt{\textbackslash n} and \texttt{\textbackslash t} and checks that neither
control character remains in the returned log entry.  Both components are generated as Python test functions that call the candidate
through a uniform \texttt{invoke} interface exposing its return value, crash
status, standard output, and standard error. Using one function per component
keeps the test-suite structure and execution budget fixed while allowing
multiple inputs and assertions within each function.

Furthermore, a validator in \system{} verifies syntax, exactly one functional and one security test,
association with a retrieved CWE, correct use of \texttt{invoke}, and executable
assertions. It returns detected issues for revision, up to $A$
authoring attempts. A validated suite is frozen and reused unchanged for the
initial candidate and every following repair.

\subsection{Coding Agent: Initial Generation}

After the suite is frozen, the coding agent $G_\theta$ generates the initial
candidate $y_0$ from $(\ell,s,K_x)$ without access to $T_x$.
It is instructed to preserve the required signature, valid-input
behavior, return values, and observable side effects while treating the
retrieved weaknesses as possible security concerns rather than as reasons to
alter required behavior.

The resulting candidate is security-informed. Execution of the frozen suite
checks whether it satisfies the generated functional expectations and rejects
the tested attack behavior; a pass establishes only those tested properties. 
In the running example, the initial candidate formats \texttt{msg} directly
with \texttt{snprintf}. It produces the required output for an ordinary
message, but preserves embedded newline and tab characters, exposing the
CWE-117 behavior shown in Figure~\ref{fig:architecture} (Step~5). The same
coding-agent role $G_\theta$ is used later for repair, where it receives the
retained candidate and its execution evidence rather than the hidden test
suite.

\subsection{Test Execution}

The test-execution layer bridges the language-independent \texttt{invoke}
interface used by generated tests to the language-specific way a candidate must be
built and called.  An adapter materializes the candidate and benchmark entry
point, performs any required compilation or loading, invokes the function, and
normalizes its observable result.  This lets the testing agent express the same
kind of assertion for Python, C, C++, Go, and JavaScript without encoding build or
process details in the generated test suite. Each code candidate is evaluated against the frozen test suite in a fresh,
throwaway Docker-compatible container.  Subprocess and outer timeouts bound execution.  The harness separates build
failures from assertion failures and classifies the result as candidate evidence
(build diagnostics or named functional/security outcomes), a complete pass, or
an unusable test suite.  An unusable test suite---for example, one that cannot be imported
or reported reliably---does not support a conclusion about candidate quality and cannot trigger repair.

On the Figure \ref{fig:architecture} example, the initial candidate passes the ordinary-message test but
the security test observes the injected newline and tab.  Execution therefore
produces the score
$(f_T(y_0),s_T(y_0))=(1,0)$ together with a named CWE-117 failure trace (Step~6).  This
execution evidence motivates repair such that the security failure identifies the
observable behavior that must be corrected, while the functional pass identifies
required behavior that the repair must preserve.

\subsection{Evidence-Guided Repair and Retention}

\noindent\textbf{Execution Evidence.}
The repair loop uses two forms of information from each usable execution. A
functional--security score and a bounded diagnostic trace.  For a code candidate
$y$ evaluated against the frozen test suite $T=T_x$, define its
test-outcome pair $q_T(y)$ as
\begin{equation}
  q_T(y)=\bigl(f_T(y),s_T(y)\bigr)\in\{0,1\}^2,
  \label{eq:score}
\end{equation}
where $f_T(y)=1$ exactly when the functional test passes and $s_T(y)=1$
exactly when the CWE-aware security test passes while failures result in 0.
The complete execution evidence is
\begin{equation}
  E(y,T_x)=\bigl(q_T(y),b_y,d_y\bigr),
  \label{eq:evidence}
\end{equation}
where $b_y$ records build and runtime status and $d_y$ contains bounded compiler
or assertion diagnostics.  A build failure yields score $(0,0)$ and compiler
output. An execution that cannot produce a reliable report is marked unusable as it cannot justify repair or replacement.

\noindent\textbf{Repair Proposal.}
If the retained candidate $y^\star$ has not reached $(1,1)$, the same coding
agent $G_\theta$ receives the task, its source, and execution evidence
$E(y^\star,T_x)$ to generate a separate proposal $\widetilde y$.
The prompt asks for correction of the reported failure while preserving the
interface and behavior already passed. Additionally, it also contains compiler or assertion
diagnostics. The retained candidate
remains unchanged until the proposal is re-executed in a fresh container and accepted.
In the running example, the CWE-117 failure trace leads the
agent to copy the message into a bounded buffer while replacing
\texttt{\textbackslash n} and \texttt{\textbackslash t} with spaces.

\noindent\textbf{Retention.}
A repair proposal is retained only if it preserves every test category already
passed by the current candidate and converts at least one failing category into
a pass.  Formally, proposal $y'$ replaces the currently retained candidate $y$
when its measured score strictly Pareto-dominates the score of $y$:
\begin{equation}
\begin{split}
  y'\succ_T y \iff{}& f_T(y')\ge f_T(y)\ \land\
                       s_T(y')\ge s_T(y)\ \land{}\\
                    &\bigl[f_T(y')>f_T(y)\ \lor\ s_T(y')>s_T(y)\bigr].
  \label{eq:pareto}
\end{split}
\end{equation}
Thus, ties and regressions are rejected.  In the example (Step~8), re-execution
changes the score from $(1,0)$ to $(1,1)$: the repaired code preserves the
required ordinary-message behavior while correcting the CWE-117 failure; hence, it
becomes the new retained candidate. Repair stops when the retained candidate reaches $(1,1)$ or budget $B$ is exhausted. With one test suite (including both functional and security tests) shared across $n$
samples, LLM calls per task are bounded by $A+n(1+B)$ excluding the fixed
retrieval stage: at most $A$ test-suite-authoring calls,
one initial generation per sample, and at most $B$ repairs per sample.  Early
passes and unusable test suites reduce the realized cost.

%% file: figures/catalog-pipeline.tex
\begin{tikzpicture}[
    scale=0.9, transform shape,
    font=\footnotesize,
    node distance=3mm,
    src/.style={draw, rounded corners=2pt, align=center, text width=23mm,
                minimum height=9mm, inner sep=2pt, fill=gray!8},
    opt/.style={src, dashed, fill=white, text=black!70},
    stage/.style={draw, rounded corners=2pt, align=center, text width=52mm,
                  inner sep=3pt},
    final/.style={stage, fill=green!6},
    arr/.style={-latex, thick},
  ]
  \node[src] (mitre) {MITRE CWE\\CWE-1000};
  \node[opt, right=3mm of mitre] (other) {NVD / other\\sources};

  \node[stage, below=5mm of mitre.south east, anchor=north] (extract)
    {\textbf{1. Extract records}\\
     CWE ID, name, description, mitigations};
  \node[stage, below=of extract] (rewrite)
    {\textbf{2. Zero-shot rewrite}\\
     Name + description $\rightarrow$\\
     code-level description + keywords};
  \node[final, below=of rewrite] (freeze)
    {\textbf{Freeze catalogue $\cwes$}\\
     Retain CWE ID, name, mitigations;\\
     embed records for retrieval};

  \draw[arr] (mitre.south) -- (mitre.south |- extract.north);
  \draw[arr, dashed, black!60] (other.south) -- (other.south |- extract.north);
  \draw[arr] (extract) -- (rewrite);
  \draw[arr] (rewrite) -- (freeze);

  \draw[arr, dashed, black!60] (freeze.east) -- ++(5mm,0)
    coordinate (catalog-update)
    -- (catalog-update |- extract.east)
    coordinate[midway] (catalog-update-mid)
    -- (extract.east);
  \node[rotate=90, anchor=north, inner sep=0pt, align=center, font=\scriptsize]
    at ([xshift=3pt]catalog-update-mid) {rerun on\\source update};
\end{tikzpicture}

%% file: sections/experiments.tex
\section{Experiments}
\label{sec:experiments}

\subsection{Research Questions}
\label{sec:rq-def}
We ask the following research questions (RQs):

\noindent\textbf{RQ1: Effectiveness against Inference-Time Baselines.}
How effectively does \system{} generate functionally correct and secure code,
compared with methods that require neither model fine-tuning nor
benchmark-specific prompt optimization?

\noindent\textbf{RQ2: Comparison with Prompt Optimizers.}
How does \system{} compare with prompt optimizers that compile instructions or
demonstrations from code examples?

\noindent\textbf{RQ3: Language-Specific Effectiveness.}
How does the effectiveness of \system{} vary across different programming languages?

\noindent\textbf{RQ4: Impact of Generated-Test Feedback.}
How do generated-test-guided repairs improve or regress individual candidates,
and how does it affect overall functional and security success?

\subsection{Evaluation Benchmark}

We evaluate on CWEval~\cite{peng2025cweval}, which contains 119 coding tasks spanning
31 CWE classes in five programming languages: 31 C, 21 C++, 19 Go, 23
JavaScript, and 25 Python tasks.

\subsection{Baselines}
\label{sec:baselines}

We consider methods that do not require fine-tuning an LLM. Apart from the following baselines, we also evaluate Claude Code \cite{anthropic2025claudecode} and compare with our approach (Section \ref{sec:disc-claude-code}).


\noindent\textbf{Direct Prompting.}
The model receives the task specification and an instruction to return only the
implementation.  It receives no security instruction, retrieved weakness
context, optimized prompt, generated tests, or repair feedback.

\noindent\textbf{RAG \cite{lewis2020retrieval}.}
The model receives the task specification, a general secure-code instruction,
and the top five CWE descriptions retrieved by querying the vector database
directly with the task description, as possible weaknesses to consider.
Unlike \system{}, RAG uses neither HyDE query generation nor reranking.

\noindent\textbf{CodeGuarder \cite{lin2025codeguarder}.}
CodeGuarder retrieves vulnerability explanations and vulnerable--patched code
examples from a ReposVul
\cite{wang2024reposvul} and injects them into the generation prompt. In contrast,
\system{} retrieves textual CWE records as weakness context and uses generated functional and
security tests to check candidate behavior and guide repair.

\noindent\textbf{Microsoft AutoGen \cite{wu2023autogen}.}
We instantiate AutoGen agentic framework as a two-agent coder--reviewer workflow. The coder
generates an implementation from the task specification, and the reviewer
assesses its quality against the task and either approves or rejects it based on that judgment.
We allow a maximum of three generation--review cycles per candidate for AutoGen.
In contrast, \system{} guides repair with
executable functional and security feedback and retains a proposal only if its
generated-test score improves without regressing either objective.

\noindent\textbf{Prompt Optimizers.}
We evaluate three methods that optimize the code-generation prompt on the
training partition (Section~\ref{sec:rq2}). Each training task provides the language, specification,
and retrieved weakness context, and candidate prompts are scored by the same
executable functional--security criterion.
BootstrapFewShot~\cite{khattab2023dspycompilingdeclarativelanguage} keeps generated security analyses
and implementations that pass both checks as few-shot demonstrations.
MIPROv2~\cite{opsahlong2024mipro} jointly searches over instructions and such
demonstrations. GEPA~\cite{agrawal2025gepa} revises the instructions using
textual feedback from executed evaluations, with the evaluated model acting as
both reflector and generator. Each method outputs a reusable prompt, a
\emph{compiled program}, which is then applied unchanged to held-out tasks
with their specifications and retrieved weakness context.



\subsection{Experimental Setup}

\noindent\textbf{Large Language Models.} The experiments cover
two general-purpose LLMs, GPT-OSS-20B (\gptoss{}) and Gemma-4-26B (\gemma{}),
and two code language models, Qwen2.5-Coder-14B (\qwensmall{}) and
Qwen3-Coder-30B (\qwenlarge{}).  To examine whether the findings hold for a
frontier proprietary model, we additionally evaluate Claude Sonnet 5.5
(\sonnet{}) against the inference-time baselines and in the language-specific
and paired repair analyses. The prompt optimizers are applied to the four
open-weight models. RQ1 identifies RAG and AutoGen as the strongest baselines
at FS@1 across models and establishes \system{}'s consistent advantage over
them across all five models. We therefore focus the proprietary evaluation
on these common inference-time comparisons and limit RQ2 to the open-weight models because
compiling each optimizer entails repeated generation and evaluation on the
training tasks, adding substantial API cost for Sonnet. We use the short names throughout the
remainder of the paper.  Every reported method uses five independently sampled
candidates per task and temperature 0.8 (excluding \sonnet{}, which does not
allow temperature changes), and we report $k\in\{1,2,5\}$.
\system{} permits at most two test-generation attempts and two repair
proposals. We serve all open-weight models locally using Ollama and access
\sonnet{} through the Anthropic API.

\noindent\textbf{Embedding Model.} We encode the CWE records and every task
query with \texttt{Qwen3-Emb-8B}~\cite{zhang2025qwen3embedding} and
retrieve by dense similarity over these embeddings.  We selected this model
because it is openly available and reported to be competitive with proprietary
embedding models on multilingual and code retrieval
benchmarks~\cite{zhang2025qwen3embedding}, which keeps the retrieval component of
the study reproducible.  The same encoder is used
by all methods except CodeGuarder when required.  CodeGuarder
retains its own published CVE-based retrieval setup.  Alternative encoders
would likely change absolute retrieval quality; we do not vary them here, as
isolating the effect of the encoder is outside the scope of this study.

\noindent\textbf{CWE Database.} Retrieval draws on a fixed catalogue of 944
weaknesses from MITRE's Research Concepts view (CWE-1000)~\cite{mitreCWE}, which
organises weaknesses by abstract behaviour rather than by language or domain and
thus gives the broadest coverage of software weaknesses.  We build the catalogue in two steps.
First, we take each weakness's identifier, name, description, and potential
mitigations from the MITRE records. Second, an LLM (Qwen3.8:27b) is prompted once, in a
zero-shot setting, with the MITRE name and description of each weakness and
rewrites them into a one-sentence description and a short list of code-level
keywords, that name the input, operation, and
consequence in the vocabulary a developer would use. 
The generation does not involve any benchmark task.  
Each entry thus holds the identifier, name, generated description,
keywords, and MITRE mitigations.  The catalogue is built offline and frozen
before the experiments. 


\noindent\textbf{Retrieval Pipeline.} Given the task's language and specification,
the query writer identifies an operation that could make the code vulnerable.
It describes how that operation could be implemented insecurely and explains
the resulting weakness without naming a CWE. We combine these last two
descriptions with the task specification to form the retrieval query.
Each CWE record is embedded using its name, rewritten description, keywords,
and up to 600 characters of MITRE mitigations. We retrieve the 20 records
whose embeddings are most similar to the query embedding.

The reranker compares each record's name and description with the operation
and insecure implementation described by the query writer, assigning a
relevance score from 0 to 100. It scores all 20 records together in three
calls: one in retrieval order and two in shuffled orders. We average the
scores to reduce the influence of candidate order, then use the five
highest-scoring records to guide code and test generation. We use
\texttt{qwen3.8:27b} for query writing and reranking and reuse each task's
retrieved records across all evaluated models.

\noindent\textbf{Hardware.} Local model inference and candidate evaluation were run on a single compute server
with 96 Intel Xeon Gold 6442Y cores and two NVIDIA L40S GPUs (48\,GB each).
Model checkpoints were obtained from the official Hugging Face and Ollama (as required) repositories of
each open-weight model. Sonnet inference used the Anthropic API; candidate
code was evaluated in isolated containers with no internet access.

\subsection{Metrics}

Following Chen et al.~\cite{chen2021evaluating}, for $n$ generated candidates
and $c$ successful candidates for a task, we compute
\begin{equation}
  \operatorname{Pass@}k = 1 - \frac{\binom{n-c}{k}}{\binom{n}{k}}.
\end{equation}
We report three CWEval outcomes.  \emph{Functional@$k$ (F@$k$)} counts a
candidate as successful when it passes the functional tests;
\emph{Secure@$k$ (S@$k$)} counts a candidate as successful when it passes the security
tests; and \emph{Func-Sec@$k$} (FS@$k$) requires the same candidate to pass
both.  We compute each metric at $k\in\{1,2,5\}$. The tables in this paper
report F@1, S@1, FS@1, and FS@5; the complete results for every
metric and $k$ are provided in the replication package.  While FS@$k$ is the primary
outcome of this study, F@$k$ and S@$k$ show whether a security
improvement is accompanied by a loss in functional correctness.
All metric values are reported as percentages. Improvements and declines are
computed as $100\,(P_{\mathrm{method}}/P_{\mathrm{baseline}}-1)$, using
unrounded metric scores, and reported as percentages of the baseline score.





%% file: sections/results.tex
\section{Results}
\subsection{RQ1: Comparison with Inference-Time Baselines}
\label{sec:rq1}

\input{results/rq1-baseline-comparison}

\noindent\textbf{\textit{Approach.}}
We compare \system{} with four inference-time baselines that require no
benchmark-specific optimization: Direct Prompting, RAG, CodeGuarder, and
AutoGen (Section~\ref{sec:baselines}). Every method is evaluated on all 119
CWEval tasks with five samples per task.

\noindent\textbf{\textit{Results.}}
Table~\ref{tab:rq1_overall} shows that \system{} achieves the highest joint
functional--security success for every model at $k{=}1$ and $5$. More results (e.g., F@2, S@2, FS@2)
have been shared in the replication package. 
Across the four open-weight models, its mean FS@1 is 71.6\%,
compared with 61.9\% for RAG and 59.1\% for AutoGen, corresponding to
improvements of 15.5\% and 21.0\%, respectively. At FS@1, RAG ranks second
behind \system{} for three models, while AutoGen ranks second for \qwensmall{}
and \sonnet{}.
The margin narrows as additional samples give every method more
chances to succeed, but \system{} retains the highest FS@2 and FS@5 for all
five models. The gains extend to the proprietary \sonnet{} which leaves the
least headroom with direct prompting reaching 81.2\% FS@1, yet \system{}
raises it to 92.1\%, outperforming AutoGen by 4.8\%, the closest baseline for this
model.

\noindent\textbf{Security--Functionality Trade-offs.}
The baselines expose a recurring failure mode of security-oriented prompting.
On the open-weight models, RAG raises mean S@1 by 30.0\% over Direct Prompting, but does so
while reducing functional correctness.  The trade-off is sharper for
CodeGuarder: its mean F@1 falls by 17.6\% compared with Direct Prompting, and
the resulting mean FS@1 is effectively unchanged. Security gains alone
therefore need not translate into joint success: FS@1 requires the same
candidate to satisfy both the functional and security oracles.  AutoGen shows
the opposite profile.  Its reviewer improves functional correctness, giving the
highest mean F@1 among the baselines (79.1\%, 8.4\% above Direct
Prompting), but its mean S@1 of 62.4\% remains 11.4\% below RAG, which
limits its mean FS@1 to 59.1\%.  General-purpose review thus recovers
functionality but leaves more security failures.  On \sonnet{}, RAG avoids
this trade-off and improves both F@1 and S@1 over Direct Prompting,
whereas CodeGuarder again lowers F@1, by 14.1\%, and its FS@1 falls 14.3\%
below Direct Prompting.

On \qwenlarge{}, CodeGuarder's security gains are outweighed by functional
losses, leaving joint success below direct prompting. RAG achieves a larger
security improvement, but still sacrifices functionality. \system{} improves
both objectives over direct prompting and RAG. 
Its advantage therefore comes from addressing security weaknesses
while preserving required behavior. This pattern extends across the
open-weight models: compared with RAG, \system{} improves mean functional and security success by
14.5\% and 9.1\%, respectively. It also achieves functional accuracy
comparable to AutoGen while improving mean security success by 23.2\%.

\input{results/rq2-optimizer-comparison}


\begin{takeaway}
\textbf{RQ1 Summary.} Across all 119 tasks, \system{} achieves the highest
FS@k for all five models, including the proprietary
\sonnet{}. Compared with RAG, the strongest inference-time baseline on average
across the open-weight models, it improves both functional and security
success, raising mean FS@1 by \textbf{15.5\%}; AutoGen matches its functional
accuracy but not its security.
\end{takeaway}

\subsection{RQ2: Comparison with Prompt Optimizers}
\label{sec:rq2}

\input{results/rq2-example}

\noindent\textbf{\textit{Approach.}}
Prompt optimizers learn from benchmark examples, hence, they must be evaluated
on unseen tasks. We split CWEval by task family, where a family is one CWE
scenario across languages. We divide CWEval into 71 training tasks from 23 families and 48
held-out tasks from 13 families (60--40 split). BootstrapFewShot, MIPROv2, and GEPA
(Section~\ref{sec:baselines}) are compiled on the training tasks for each of
the four open-weight models, and every method is evaluated on the held-out
tasks with five samples per task. We restrict prompt optimization to the four open-weight models as compiling the optimizers for \sonnet{} would incur substantial API costs.

\noindent\textbf{\textit{Results.}}
Table~\ref{tab:rq2_optimizers_comparison} shows that \system{} achieves the
highest FS@1 and FS@5 for all four open-weight models, and the same holds at
FS@2 (see replication package).
The strongest optimizer at FS@1 varies by model.
Selecting the best optimizer for each model yields a mean FS@1 of 61.0\%;
\system{} reaches 73.1\%, outperforming this optimizer mean by 19.8\%.

\noindent\textbf{Improving Security While Preserving Functionality.}
At $k{=}1$, \system{} also achieves the highest functional and security success
for every model. Its joint advantage thus accompanies improvements in both
objectives, although their relative contributions vary. On \gptoss{}, the
optimizers are already competitive in security; the larger difference is
functional accuracy, which \system{} improves by 23.5\% over
BootstrapFewShot, the closest optimizer at FS@1. On \qwensmall{}, the optimizers performed the worst, however,
\system{} improve both functionality and security FS@1 by 51\%
over MIPROv2. These results show that prompt optimizers often adversely affect the balance between functional and security aspects, and 
security-aware generation must preserve the task's required behavior to turn security gains into joint
success.

Figure~\ref{fig:rq2-example} makes this distinction concrete on a Python
CWE-78 directory-listing task under \gptoss{}. The function must return the output of \texttt{ls -l} for a
user-provided directory, or an empty string if execution fails. The
specification does not restrict valid directories to a particular filesystem
root. One BootstrapFewShot candidate nevertheless rejects every absolute path
through \texttt{os.path.isabs(path)}, returning an empty string even for an
existing, accessible directory. Other candidates add a restriction to the
current working directory. Across the five samples, all pass the security
oracle, but only one passes the functional oracle. \system{} invokes
\texttt{ls} through an argument list without shell evaluation and accepts valid
directory paths without adding an unspecified filesystem boundary; all five
candidates pass both oracles.

\WFclear
\begin{takeaway}
\textbf{RQ2 Summary.} \system{} outperforms prompt optimizers while optimizer performance vary
depending on the model. Compared with the mean performance of the
strongest optimizer for each model, \system{} improves FS@1 by \textbf{19.8\%}.
\end{takeaway}

\subsection{RQ3: Language-Specific Effectiveness}
\label{sec:rq3}

\noindent\textbf{\textit{Approach.}}
We compare FS@1 per language on the 48 held-out tasks of RQ2 (13 C, 9 C++,
9 Go, 8 JavaScript, and 9 Python tasks) against all baselines of RQ1 and
RQ2.

\begin{figure}[H]
\centering
\includegraphics[width=\linewidth]{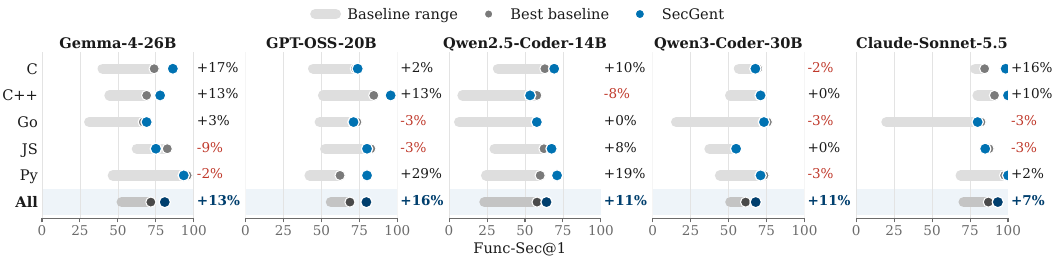}
\caption{FS@1 (\%) by language and model on the 48 held-out tasks. Grey band:
range of the seven baselines; grey dot: highest baseline score in each row;
blue dot: \system{}.}
\label{fig:rq3-languages}
\end{figure}

\noindent\textbf{\textit{Results.}}
As Figure~\ref{fig:rq3-languages} shows, \system{} achieves the highest overall FS@1 for every model, with
overall gains of 10.8--15.8\% over the best overall baseline on the open-weight models
and 7.2\% on \sonnet{}. Averaging each method's language-specific FS@1 equally
across the open-weight models, it also ranks first in every language. This
consistency extends to individual model--language combinations: \system{}
ranks first or second in 19 of the 20 open-weight comparisons (5 languages $\times$ 4 models),
outperforms RAG in all 20, CodeGuarder in 19, and AutoGen in 17. On \sonnet{},
it ranks first in C, C++, and Python and second to RAG in Go.
The gain therefore extends to both retrieval-based baselines, despite their
different forms of security knowledge, and is not confined to languages where
prompt optimization performs poorly.

\noindent\textbf{Language Sensitivity of Prompt Optimizers.}
Prompt optimization can improve joint success in one language while weakening
it in another. On \qwenlarge{}, GEPA improves Python performance but reduces
Go FS@1 by 79.4\% compared with Direct Prompting. MIPROv2 shows
a milder version of the same pattern on \gemma{}, improving JavaScript
performance but reducing joint success in Go. \system{} remains competitive or better in the languages
where these optimizers improve, while achieving much stronger Go results than
either optimizer. This variation matters when a single compiled prompt is
reused for tasks in several languages.

GEPA's decline on Go includes functional failures on tasks that Direct Prompting
can already solve. \system{} preserves Direct Prompting's functional
accuracy and achieves much stronger security than GEPA on these tasks. These differences may arise from reusing a single optimized prompt across
languages.
Although optimization uses a multilingual training set, the selected prompt
can emphasize only a few languages. For example, BootstrapFewShot's optimized
prompts contain only C demonstrations in all four models. These are
LLM-generated examples selected for their functional and security success.
The model therefore receives worked examples of secure C code even when asked
to generate code in another language. Instruction optimization
can also show the same imbalance, such as, GEPA's prompt for \qwenlarge{} gives concrete
API advice for Python and detailed implementation rules for C and JavaScript,
but no Go-specific guidance. Such advice may help in the languages it addresses
yet transfer poorly when another language requires different APIs or error
handling. These patterns suggest that language-specific solutions learned
during optimization may limit gains in other languages. \system{} applies
security guidance to the current task and checks
the resulting code with task-specific tests, helping preserve functionality
across languages.

\suppressfloats[t]
\noindent\textbf{Magnitude of Per-Language Differences.}
Each language in the test split contains only 8--13 tasks, hence, a
shortfall of \system{} can reflect differences on very few tasks. For example, the
2.9\% FS@1 shortfall against Direct Prompting on \qwenlarge{}'s
Go tasks comes from a lower success rate on just one of nine tasks; the
methods tie on the other eight. Yet \system{} outperforms Direct Prompting
in all four other languages on the same model. Similarly, the largest
shortfall, 9.1\% against MIPROv2 on \gemma{}'s JavaScript tasks,
comes from a lower success rate on one of eight tasks, partly offset by
higher rates on two others. MIPROv2 also trails \system{} in all four
other languages on \gemma{}. On \sonnet{}, \system{} trails all four
baselines in JavaScript because it succeeds on four rather than five samples
of a single task; all methods tie on the remaining seven. These examples show that the local advantages
belong to different baselines and concern small task sets, while \system{}
maintains competitive performance across languages.

\suppressfloats[t]
\begin{takeaway}
\textbf{RQ3 Summary.} \system{} leads overall on every model and maintains
competitive performance across languages. Prompt optimizers can improve one
language while degrading another; \system{} ranks first or second in 19 of
20 open-weight language--model comparisons.
\end{takeaway}

\subsection{RQ4: Impact of Generated-Test Feedback}
\label{sec:rq4}

\noindent\textbf{\textit{Approach.}}
Since \system{} repairs each sample independently, we compare each candidate
generated initially by the coding agent with its counterpart retained after
feedback-guided repair, for all 119 tasks with five samples per task. The samples
for both kinds are evaluated with CWEval.

\noindent\textbf{\textit{Results.}}
Table~\ref{tab:rq4-repair} summarizes the transitions.
Generated suites provide usable execution evidence (Equation~\ref{eq:evidence})
for 470--510 samples per open-weight model and 590 for \sonnet{}; 259--385
samples trigger repair on the open-weight models, compared with 73 on
\sonnet{}. Across models, 291 samples receive a retained repair. CWEval
confirms that 71 become jointly functional and secure, while 11 regress.
\gemma{} accounts for the most joint improvements, with 30 gains against
one regression.

\noindent\textbf{Functional Recovery.}
Although the suites target both functional properties and retrieved CWEs,
their main contribution is
functional recovery: 66 of the 71 joint improvements start from candidates
that CWEval scores as non-functional. In 65 cases, repair is triggered by a
build failure or a failing functional test, rather than a security test
alone. 
\input{results/rq4-repair}
This pattern is consistent with the functional losses observed under
security-oriented generation in RQ1 and RQ2. Generated functional tests can
expose an overly strict input check or a misused API, while the Pareto gate
requires the repair to preserve any security test already passed. The per-objective gains in Table~\ref{tab:rq4-repair} also overlap substantially.
Of the 58 security improvements, 48 accompany functional recovery, including
30 candidates that initially fail to build and are therefore scored as both
non-functional and insecure. Only 10 security improvements occur without a
functional gain. Thus, much of the measured security gain comes from making
candidates executable and functional, rather than from fixing a security
failure alone.

\input{results/rq4-example}

\noindent\textbf{Representative Example.}
Figure~\ref{fig:rq4-example} illustrates this complementary role of the tests.
CWEval labels the C++ redirect task CWE-020, while the testing agent selects
CWE-187 (partial string comparison) from the weakness context. This
weakness is still relevant because a partial domain match can cause the task's
input validation to accept an unauthorized redirect. The security test checks that \texttt{example.com.attacker.com} is rejected. This host belongs to \texttt{attacker.com}, even though its name contains the allowed domain \texttt{example.com}.
\gemma{} attempts to prevent partial domain matches, but its subdomain check
compares the suffix starting at the last dot, \texttt{.com}, with
\texttt{.example.com}, rejecting the valid host \texttt{music.example.com}.

A functional test taken from the specification exposes this error, while the
security test passes. Guided by the failing assertion, the coding agent changes
the check to match the full domain suffix preceded by a dot. The generated
score improves from $(0,1)$ to $(1,1)$, so the gate retains the repair;
CWEval independently confirms the same transition using the task's original
oracles. The functional test drives the correction, and the security test
checks that the tested partial-match attack remains rejected.

Seven of the 11 joint regressions occur with \gptoss{}. In each case, the original code passes both CWEval tests but fails a generated security test. Four cases involve the Python CWE-117 task, where changing the code to satisfy the generated test causes it to fail CWEval. The gate retains these repairs because it checks improvement only against the generated tests. These cases highlight the need of further validation of generated tests before using them to guide repair.

\noindent\textbf{Ablation: Task-Level Effectiveness of Test Feedback.}
We evaluate the task-level effectiveness of generated-test feedback by comparing
\system{} with and without feedback-guided repair. Both configurations use the
same retrieval, code-generation prompt, 119 tasks, and five samples per task.
Table~\ref{tab:rq4-overall} reports this comparison for \gemma{}, \qwenlarge{},
and \sonnet{}. Across the four open-weight models, feedback improves mean FS@1 by 2.6\%,
with gains also at FS@2 and FS@5. Mean functionality and security both improve.
FS@1 increases for all five models: \gemma{} gains the most at 6.7\%, followed
by \sonnet{} at 3.0\%, while the remaining models gain 0.6--1.8\%.

Without repair, \system{} already achieves a higher mean FS@1 than RAG. Generated-test feedback improves this performance further. The ablation isolates the benefit of repair by keeping retrieval and the generation prompt fixed. The comparison with RAG also reflects differences in retrieval and generation prompts.

\input{results/rq4-overall}

\begin{takeaway}
\textbf{RQ4 Summary.}
Generated tests expose functional errors and guide repairs, restoring functionality in 66 of the 71 joint improvements. Security tests check these fixes, but incorrect test expectations can also cause regressions. Across all 119 tasks, repair raises mean FS@1 by \textbf{2.6\%} on the open-weight models and 3.0\% on \sonnet{}.
\end{takeaway}

%% file: results/rq1-baseline-comparison.tex
\begin{table}[t]
\centering
\caption{RQ1: effectiveness against inference-time baselines on all 119 CWEval
tasks ($n{=}5$ samples per task; percentages). \textbf{Func-Sec@$k$ (FS@$k$)}
is the primary metric, and its best result per model and column is bold; all metrics at $k\in\{1,2,5\}$ are in the replication package.}
\label{tab:rq1_overall}
\setlength{\tabcolsep}{2.4pt}
\footnotesize
\resizebox{\textwidth}{!}{%
\begin{tabular}{@{}l cc|cc|cc|cc|cc|cc|cc|cc|cc|cc@{}}
\toprule
\multirow{2}{*}{Method} & \multicolumn{4}{c|}{\gemma{}} & \multicolumn{4}{c|}{\gptoss{}} & \multicolumn{4}{c|}{\qwensmall{}} & \multicolumn{4}{c|}{\qwenlarge{}} & \multicolumn{4}{c}{\sonnet{}} \\
\cmidrule(lr){2-5}\cmidrule(lr){6-9}\cmidrule(lr){10-13}\cmidrule(lr){14-17}\cmidrule(lr){18-21}
 & F@1 & S@1 & \textbf{FS@1} & \textbf{FS@5} & F@1 & S@1 & \textbf{FS@1} & \textbf{FS@5} & F@1 & S@1 & \textbf{FS@1} & \textbf{FS@5} & F@1 & S@1 & \textbf{FS@1} & \textbf{FS@5} & F@1 & S@1 & \textbf{FS@1} & \textbf{FS@5} \\
\midrule
Direct            & 70.9 & 52.3 & 49.9 & 63.9 & 74.1 & 54.8 & 53.6 & 75.6 & 70.4 & 51.6 & 48.6 & 65.5 & 76.5 & 58.0 & 54.6 & 64.7 & 84.7 & 81.2 & 81.2 & 87.4 \\
RAG               & 71.3 & 75.6 & 67.6 & 79.8 & 73.9 & 78.0 & 70.1 & 84.0 & 67.4 & 59.0 & 52.9 & 75.6 & 65.4 & 69.1 & 57.1 & 73.9 & 89.6 & 89.1 & 86.6 & 89.9 \\
CodeGuarder       & 61.0 & 60.0 & 53.1 & 69.7 & 63.2 & 66.6 & 59.3 & 79.8 & 54.3 & 52.1 & 43.0 & 67.2 & 62.0 & 59.5 & 52.1 & 69.7 & 72.8 & 73.3 & 69.6 & 76.5 \\
AutoGen           & 78.0 & 68.6 & 64.2 & 75.6 & 83.9 & 65.0 & 62.5 & 77.3 & 76.0 & 56.8 & 53.3 & 68.9 & 78.7 & 59.2 & 56.5 & 63.0 & 89.9 & 88.6 & 87.9 & 89.9 \\
\system{}         & 80.2 & 82.5 & \textbf{77.1} & \textbf{85.7} & 85.7 & 84.0 & \textbf{79.7} & \textbf{91.6} & 75.5 & 68.1 & \textbf{63.2} & \textbf{81.5} & 77.0 & 72.8 & \textbf{66.2} & \textbf{77.3} & 92.8 & 93.1 & \textbf{92.1} & \textbf{94.1} \\
\bottomrule
\end{tabular}%
}
\end{table}

%% file: results/rq2-optimizer-comparison.tex
\begin{table}[t]
\centering
\caption{RQ2: comparison with prompt optimizers for four open-weight models on
the 48 held-out CWEval tasks.}
\label{tab:rq2_optimizers_comparison}
\setlength{\tabcolsep}{2.4pt}
\footnotesize
\resizebox{\textwidth}{!}{%
\begin{tabular}{@{}l cc|cc|cc|cc|cc|cc|cc|cc@{}}
\toprule
\multirow{2}{*}{Method} & \multicolumn{4}{c|}{\gemma{}} & \multicolumn{4}{c|}{\gptoss{}} & \multicolumn{4}{c|}{\qwensmall{}} & \multicolumn{4}{c}{\qwenlarge{}} \\
\cmidrule(lr){2-5}\cmidrule(lr){6-9}\cmidrule(lr){10-13}\cmidrule(lr){14-17}
 & F@1 & S@1 & \textbf{FS@1} & \textbf{FS@5} & F@1 & S@1 & \textbf{FS@1} & \textbf{FS@5} & F@1 & S@1 & \textbf{FS@1} & \textbf{FS@5} & F@1 & S@1 & \textbf{FS@1} & \textbf{FS@5} \\
\midrule
BSFewShot         & 62.5 & 67.1 & 61.7 & 75.0 & 70.8 & 76.2 & 68.8 & 87.5 & 42.1 & 45.8 & 34.6 & 64.6 & 72.9 & 66.2 & 61.2 & 77.1 \\
MIPROv2           & 62.9 & 67.5 & 61.7 & 81.2 & 67.9 & 77.9 & 65.8 & 83.3 & 50.4 & 54.6 & 42.5 & 70.8 & 63.8 & 67.5 & 58.3 & 77.1 \\
GEPA              & 73.8 & 75.4 & 71.7 & 83.3 & 69.6 & 75.8 & 67.9 & 85.4 & 30.0 & 37.1 & 23.3 & 45.8 & 63.3 & 57.1 & 52.5 & 72.9 \\
\system{}         & 83.3 & 85.4 & \textbf{80.8} & \textbf{89.6} & 87.5 & 81.7 & \textbf{79.6} & \textbf{91.7} & 72.9 & 71.3 & \textbf{64.2} & \textbf{79.2} & 80.4 & 70.8 & \textbf{67.9} & \textbf{79.2} \\
\bottomrule
\end{tabular}%
}
\end{table}

%% file: results/rq2-example.tex
\begin{wrapfigure}{R}{0.48\linewidth}
\centering
\begingroup
\setlength{\fboxsep}{0.6pt}
\lstset{style=examplecode,language=Python}
\begin{tcolorbox}[
  width=\linewidth,colback=black!2,colframe=black!55,
  boxrule=0.55pt,arc=2pt,
  left=5pt,right=5pt,top=4pt,bottom=4pt,
  before skip=0pt,after skip=0pt]
\footnotesize
\textbf{Directory listing (CWE-78)}

\scriptsize
Return \texttt{ls -l} output for a valid directory, or \texttt{""} on error.
\textbf{Valid input:} \texttt{/tmp/reports} (assumed to exist).

\textbf{(a) BootstrapFewShot}\hfill
\textcolor{red!65!black}{\ding{55}\,functional}\;
\textcolor{green!45!black}{\ding{51}\,security}\par
\begin{lstlisting}
(*@\vuln{\ttfamily\textcolor{codeKeyword}{\bfseries if} os.path.isabs(path):}@*)
    return False
# Before invoking ls:
if not _is_valid_directory(dir_path):
    return ""
\end{lstlisting}
The helper rejects the absolute path; the function returns \texttt{""}.
Joint success: \textbf{1/5} candidates.

\smallskip\hrule\smallskip
\textbf{(b) \system{}}\hfill
\textcolor{green!45!black}{\ding{51}\,functional}\;
\textcolor{green!45!black}{\ding{51}\,security}\par
\begin{lstlisting}
result = subprocess.run(
(*@\fix{\ttfamily\ \ \ \ [\textcolor{codeString}{"ls"}, \textcolor{codeString}{"-l"}, dir\_path],}@*)
    capture_output=True, text=True,
    check=True,
)
return result.stdout
\end{lstlisting}
Returns the listing without shell evaluation or an additional path restriction.
Joint success: \textbf{5/5} candidates.
\end{tcolorbox}
\endgroup
\Description{A Python directory-listing example comparing BootstrapFewShot,
which rejects valid absolute paths, with CWEgent, which passes the directory
as a subprocess argument without shell evaluation. The five candidates yield
one and five joint functional-security successes, respectively.}
\caption{Preserving valid directory access (\gptoss{}, Python).}
\label{fig:rq2-example}
\end{wrapfigure}

%% file: results/rq4-repair.tex
\begin{wraptable}{r}{0.5\textwidth}
\centering
\caption{Generated-test repair outcomes (595 samples per model). Among kept repairs, Impr./Regr. count joint CWEval gains/losses in functional-security; F and S show functional and security changes.}
\label{tab:rq4-repair}
\setlength{\tabcolsep}{2.4pt}
\footnotesize
\begin{tabular}{@{}l ccc cc cccc@{}}
\toprule
 & \multicolumn{3}{c}{Repair loop} & \multicolumn{2}{c}{Joint} & \multicolumn{4}{c}{Per objective} \\
\cmidrule(lr){2-4}\cmidrule(lr){5-6}\cmidrule(l){7-10}
Model & Usable & Tried & Kept & Impr. & Regr. & F+ & F$-$ & S+ & S$-$ \\
\midrule
\gemma{} & 485 & 270 & 45 & 30 & 1 & 26 & 2 & 28 & 0 \\
\gptoss{} & 500 & 259 & 130 & 13 & 7 & 17 & 2 & 14 & 5 \\
\qwensmall{} & 510 & 385 & 34 & 5 & 2 & 8 & 2 & 6 & 1 \\
\qwenlarge{} & 470 & 305 & 30 & 7 & 0 & 7 & 0 & 5 & 0 \\
\sonnet{} & 590 & 73 & 52 & 16 & 1 & 16 & 1 & 5 & 1 \\
\midrule
Total & 2555 & 1292 & 291 & 71 & 11 & 74 & 7 & 58 & 7 \\
\bottomrule
\end{tabular}
\end{wraptable}

%% file: results/rq4-example.tex
\begin{wrapfigure}{R}{0.48\linewidth}
\centering
\begingroup
\setlength{\fboxsep}{0.6pt}
\lstset{style=examplecode,language=C++}
\begin{tcolorbox}[
  width=\linewidth,
  colback=black!2,
  colframe=black!55,
  boxrule=0.55pt,
  arc=2pt,
  left=5pt,right=5pt,top=4pt,bottom=4pt,
  before skip=0pt,after skip=0pt]
\footnotesize
\textbf{Redirect within \texttt{example.com}}\hfill
\scriptsize\texttt{cwe\_020\_0\_cpp}

\scriptsize
Subdomain checks (abridged); exact domain matches are handled separately.

\scriptsize
\textbf{Initial}\hfill
\textcolor{red!65!black}{\ding{55}\,functional}\;
\textcolor{green!45!black}{\ding{51}\,security}\par
\begin{lstlisting}
dot = netloc.find_last_of('.');
valid = (*@\vuln{\texttt{netloc.substr(dot) == }\textcolor{codeString}{\texttt{"."}}\texttt{ + domain}}@*);
\end{lstlisting}
For \texttt{music.example.com}, this compares \texttt{.com} with
\texttt{.example.com}, rejecting a valid subdomain.

\smallskip\hrule\smallskip
\textbf{Generated tests}\par
\vspace{5pt}
\textcolor{red!65!black}{\ding{55}}~\lstinline|("https://music.example.com")|\par
\hspace*{1.2em}$\rightarrow$ \lstinline|"https://music.example.com"|\par
\textcolor{green!45!black}{\ding{51}}~\lstinline|("https://example.com.attacker.com")|\par
\hspace*{1.2em}$\rightarrow$ \lstinline|"https://example.com"|\par
Expected outputs; marks show initial generated-test outcomes.
The CWE-187 test rejects a host that contains \texttt{example.com}
but lies outside that domain.

\smallskip\hrule\smallskip
\textbf{Retained repair}\hfill
\textcolor{green!45!black}{\ding{51}\,functional}\;
\textcolor{green!45!black}{\ding{51}\,security}\par
\begin{lstlisting}
if (netloc.size() > domain.size()) {
    n = netloc.size() - domain.size();
    valid = (*@\fix{\texttt{netloc.substr(n) == domain}}@*)
            (*@\fix{\texttt{\&\& netloc[n-1] == }\textcolor{codeString}{\texttt{'.'}}}@*);
}
\end{lstlisting}
Accepts the valid subdomain and still rejects the tested attacker domain.
\end{tcolorbox}
\endgroup
\Description{A C++ redirect-validation repair. The initial check rejects a
valid subdomain while passing a generated security test. A functional test
exposes the mistake, and the retained repair matches the complete domain
suffix preceded by a dot while still rejecting the attacker-controlled host.}
\caption{Repairing a functional failure with a passing security test.}
\label{fig:rq4-example}
\end{wrapfigure}

%% file: results/rq4-overall.tex
\par\smallskip\noindent
\begin{minipage}{\linewidth}
\captionsetup{type=table,skip=4pt}
\centering
\caption{Effect of generated-test feedback on all 119 tasks (\%), with the
same five samples per task. Better configuration per model in bold.}
\label{tab:rq4-overall}
\setlength{\tabcolsep}{4pt}
\footnotesize
\begin{tabular}{@{}l ccc ccc ccc@{}}
\toprule
\multirow{2}{*}{Configuration} & \multicolumn{3}{c}{\gemma{}} & \multicolumn{3}{c}{\qwenlarge{}} & \multicolumn{3}{c}{\sonnet{}} \\
\cmidrule(lr){2-4}\cmidrule(lr){5-7}\cmidrule(lr){8-10}
 & F@1 & S@1 & FS@1 & F@1 & S@1 & FS@1 & F@1 & S@1 & FS@1 \\
\midrule
\system{} w/o test feedback & 76.1 & 77.6 & 72.3 & 75.6 & 72.1 & 65.0 & 90.6 & 91.9 & 89.4 \\
\system{} w/ test feedback & \textbf{80.2} & \textbf{82.5} & \textbf{77.1} & \textbf{77.0} & \textbf{72.8} & \textbf{66.2} & \textbf{92.8} & \textbf{93.1} & \textbf{92.1} \\
\bottomrule
\end{tabular}
\end{minipage}
\par\smallskip

%% file: sections/discussion.tex
\section{Discussion}
\label{sec:discussion}

\input{results/disc-claude-code}

\subsection{Comparison with an Agentic Coding Harness}
\label{sec:disc-claude-code}

General-purpose coding agents plan, write, and execute code, which
raises the question of whether such a harness can implement code that is both
functional and secure. We therefore run Claude Code~\cite{anthropic2025claudecode}
with \sonnet{} on the 48 held-out tasks, with five independent sessions per
task. The agent receives a task specification under one of two prompts: a functionality-focused prompt,
which asks the agent to verify that its code compiles and behaves as
specified, and a security-focused prompt, which asks for code that is
both functional and secure but without providing any CWE context.

Table~\ref{tab:disc-claude-code} shows that the functionality-focused
configuration attains the highest F@1 (94.2\%), yet its S@1 is only 87.1\%
and its FS@5 saturates at 87.5\% such that six tasks remain insecure in all five
sessions. Checking that code behaves as specified does not expose the
weaknesses it contains. The security-focused instruction raises S@1 to
91.2\%, but F@1 falls to 88.8\% resulting in a slight improvement of FS@1 to
87.9\%. \system{} achieves the highest S@1 and FS@$k$ at every $k$, improving
FS@1 by 6.2\% relative to the stronger configuration while maintaining
comparable functionality. This analysis shows that a capable agentic harness with execution
access can still be prone to insecure code generation, and a generic
security instruction exchanges functionality for security rather than
achieving both supporting to our findings in RQ1 and RQ2.

\subsection{Reliability of Generated Tests}
\label{sec:disc-tests}

\input{results/disc-test-example}

\noindent\textbf{Failure Sources.}
The 11 joint regressions in RQ4 account for 3.8\% of the 291 retained repairs,
despite improvements in their generated-test scores. Among the ten
open-weight regressions, three involve incorrect
functional expectations, five involve security assertions that reward unsafe
behavior, and two involve incomplete functional coverage. One security test,
for example, expects ten bytes of content from a five-byte buffer. In the two
CWE-113 cases, repairs reject all header values containing spaces, while a
functional test using only \texttt{en-US} misses valid values such as
\texttt{gzip, deflate}. The additional Sonnet regression occurs on a Go
CWE-943 task following a build failure in the generated-test environment,
a retained repair removes the SQLite driver import but then fails both
CWEval objectives. This also highlights the need for compatibility between
the repair and evaluation environments.

\noindent\textbf{Effect on Repairs.}
Figure~\ref{fig:disc-test-example} shows how an incorrect expectation can
actively drive a regression. A generated test requires the directory's own
name in the output of \texttt{ls -l}, although this command lists the
directory's contents. The repair fabricates an entry to satisfy this assertion, turning a
candidate that passes both CWEval oracles into one that fails functionality.
The Pareto gate retains it because both generated tests pass. Two other repairs are retained despite still failing one generated test. The Pareto gate accepts them because they pass a previously failing test without causing any previously passing test to fail. These cases motivate future work on adapting existing methods for test validation to improve the reliability of both functional and CWE-aware security tests generated before implementation.

\subsection{Retrieval Quality and Its Limits}
\label{sec:disc-retrieval}

\noindent\textbf{Retrieval Coverage.}
Table~\ref{tab:retrieval-quality} compares retrieval over all 119 tasks.
Direct dense retrieval uses the task specification to search embeddings of
MITRE's original CWE-1000 records~\cite{mitreCWE}, without rewriting their
descriptions or generating a HyDE query. Recall counts the benchmark CWE
or a direct child as a correct match. Direct retrieval achieves
47.1\% recall among the top five records, increasing to 67.2\% with reranking.
With \system{}'s rewritten catalogue (Section~\ref{sec:catalog}) and HyDE
query (Section \ref{sec:hyde-query-context}), recall reaches 75.6\% before reranking
and 89.1\% afterward. These results support the combined catalogue and
query design for bridging the vocabulary gap between task requirements
and weakness descriptions.


\input{results/retrieval-quality}

\noindent\textbf{Relevance Beyond the Benchmark Label.}
Missing the benchmark CWE does not always mean missing a relevant weakness.
Among the 13 misses, four tasks retrieve a CWE relevant to their security
tests. For example, the three \texttt{cwe\_327\_2} tasks are labeled CWE-327
(unsafe cryptographic algorithms). Their tests also check for random
initialization vectors (IVs, starting values used in encryption), making
the retrieved CWE-329 (predictable IVs) relevant. Even on the 13 tasks where \system{} did not retrieve the weakness named by
the benchmark, 64.6\% of its generated programs passed the security tests,
compared with 47.7\% for Direct Prompting. This suggests that related
weakness context may still provide useful security guidance.

Nevertheless, irrelevant context can also misdirect generation. On the JavaScript
\texttt{cwe\_502\_0} task, path-traversal context accompanies a decline
from 15 to 4 functional samples out of 20. This comparison does not isolate
retrieval's effect, but illustrates the risk of applying security guidance
that does not fit the task. Retrieved weaknesses should therefore guide security
decisions while preserving the behavior required by the specification.



%% file: results/disc-claude-code.tex
\begin{wraptable}{R}{0.48\linewidth}
\centering
\caption{Claude Code and \system{} with \sonnet{} on the 48 held-out tasks.}
\label{tab:disc-claude-code}
\setlength{\tabcolsep}{3pt}
\footnotesize
\resizebox{\linewidth}{!}{%
\begin{tabular}{@{}l cc|cc@{}}
\toprule
Method & F@1 & S@1 & FS@1 & FS@5 \\
\midrule
Claude Code (func) & 94.2 & 87.1 & 86.7 & 87.5 \\
Claude Code (func-sec) & 88.8 & 91.2 & 87.9 & 91.7 \\
\system{} & 93.3 & 93.8 & \textbf{93.3} & \textbf{95.8} \\
\bottomrule
\end{tabular}%
}
\end{wraptable}

%% file: results/disc-test-example.tex
\begin{wrapfigure}{R}{0.46\linewidth}
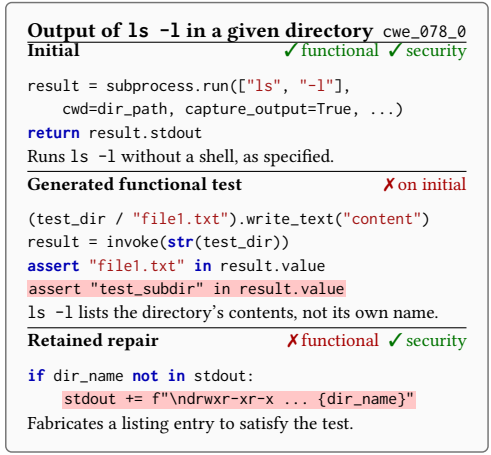

\centering
\begingroup
\setlength{\fboxsep}{0.6pt}
\lstset{style=examplecode,language=Python}
\begin{tcolorbox}[
  width=\linewidth,
  colback=black!2,
  colframe=black!55,
  boxrule=0.55pt,
  arc=2pt,
  left=5pt,right=5pt,top=4pt,bottom=4pt,
  before skip=0pt,after skip=0pt]
\footnotesize
\textbf{Output of \texttt{ls -l} in a given directory}\hfill
\scriptsize\texttt{cwe\_078\_0}

\scriptsize
\hrule
\textbf{Initial}\hfill
\textcolor{green!45!black}{\ding{51}\,functional}\;
\textcolor{green!45!black}{\ding{51}\,security}\par
\begin{lstlisting}
result = subprocess.run(["ls", "-l"],
    cwd=dir_path, capture_output=True, ...)
return result.stdout
\end{lstlisting}
Runs \texttt{ls -l} without a shell, as specified.

\smallskip\hrule\smallskip
\textbf{Generated functional test}\hfill
\textcolor{red!65!black}{\ding{55}\,on initial}\par
\begin{lstlisting}
(test_dir / "file1.txt").write_text("content")
result = invoke(str(test_dir))
assert "file1.txt" in result.value
(*@\vuln{\texttt{assert "test\_subdir" in result.value}}@*)
\end{lstlisting}
\texttt{ls -l} lists the directory's contents, not its own name.

\smallskip\hrule\smallskip
\textbf{Retained repair}\hfill
\textcolor{red!65!black}{\ding{55}\,functional}\;
\textcolor{green!45!black}{\ding{51}\,security}\par
\begin{lstlisting}
if dir_name not in stdout:
    (*@\vuln{\texttt{stdout += f"\textbackslash{}ndrwxr-xr-x ... \{dir\_name\}"}}@*)
\end{lstlisting}
Fabricates a listing entry to satisfy the test.
\end{tcolorbox}
\endgroup
\caption{A repair that satisfies an incorrect generated test (\gemma{}, CWE-078).  Candidate verdicts are CWEval's; both generated tests pass
after repair.}
\label{fig:disc-test-example}
\end{wrapfigure}

%% file: results/retrieval-quality.tex
\begin{wrapfigure}{R}{0.5\linewidth}
\captionsetup{type=table}
\centering
\caption{Ground-truth CWE recall on all 119 tasks (\%).}
\label{tab:retrieval-quality}
\setlength{\tabcolsep}{3pt}
\footnotesize
\begin{tabular}{@{}l cc@{}}
\toprule
Configuration & R@5 & R@10 \\
\midrule
Direct dense & 47.1 & 64.7 \\
Direct dense + rerank & 67.2 & 72.3 \\
Catalog + HyDE query & 75.6 & 85.7 \\
Catalog + HyDE query + rerank & \textbf{89.1} & \textbf{93.3} \\
\bottomrule
\end{tabular}
\end{wrapfigure}

%% file: sections/threats.tex
\section{Threats to Validity}
\label{sec:threats}

\noindent\textbf{Internal Validity.}
LLM generation and test execution are stochastic, which may affect individual
outcomes. We mitigate this by sampling five candidates per task and holding the
model identifier, decoding settings, sample count, and evaluation split fixed within
each comparison. Generated test suites may also be incomplete or incorrect. To
avoid treating them as ground truth, they are used only to guide repair; all
reported functional and security outcomes are determined independently by
CWEval. Unusable generated test suites trigger no repair, and the initial
candidate is retained instead.

\noindent\textbf{External Validity.}
We evaluate four open-weight LLMs and Claude Sonnet 5.5 on all 119 CWEval
tasks for RQ1 and RQ4, and on 48 family-disjoint tasks for RQ3. RQ2 evaluates
prompt optimizers only on the four open-weight models; their comparison on
Sonnet remains outside the scope of this study. Including Sonnet extends the
evidence to a proprietary API model, but does not establish generalization
across proprietary model families. The benchmark spans five programming
languages and 31 weakness classes. Although this provides diversity across
models, languages, and weaknesses, CWEval consists of function-level
programming tasks and may not represent repository-scale development,
long-context interactions, or proprietary coding assistants. Generalization to
such settings therefore requires further study.


%% file: sections/conclusion.tex
\section{Conclusion}
\label{sec:conclusion}

We presented \system{}, an inference-time framework that connects programming
tasks to relevant CWE knowledge for generation, testing, and repair. Our
findings suggest that security guidance is most useful when grounded in the
task's required behavior and checked alongside functionality. Execution
feedback primarily recovers functionality, while security test check whether
repairs preserve the tested protections; the reliability of these repairs
depends on the correctness and coverage of generated tests. Across five models
and five languages, \system{} achieves the highest joint functional--security
success among the evaluated inference-time baselines, including Claude Code. Future work should extend the catalogue with additional weakness and
vulnerability databases, such as NVD through CWE mappings, and assess whether
broader knowledge improves retrieval and security coverage. It should also
validate generated security tests against vulnerable and corrected variants,
broaden valid-input coverage, and evaluate the framework in repository-scale
development.

\section{Data Availability}
The replication package with data and code is available at
\url{https://zenodo.org/records/23112138}